\documentclass[12pt]{article}
\usepackage[margin=1in]{geometry}
\usepackage{hyperref}
\usepackage{bbding}

\usepackage[english]{babel}
\usepackage{amsmath,amsthm}
\usepackage{amssymb}
\usepackage{amsfonts}

\usepackage{algorithm}  
\usepackage{algorithmicx}  
\usepackage{algpseudocode}  

\usepackage{color}

\usepackage{CJKutf8}
\usepackage{graphicx}
\usepackage{pgf,tikz}
\usetikzlibrary{shapes,arrows,automata}

\newtheorem{thm}{Theorem}[section]

\newtheorem{lem}[thm]{Lemma}
\newtheorem{conj}[thm]{Conjecture}
\newtheorem{ques}[thm]{Question}

\theoremstyle{definition}
\newtheorem{defn}[thm]{Definition}
\newtheorem{rem}[thm]{Remark}

\newtheorem{fact}[thm]{Fact}
\numberwithin{equation}{section}

\newcommand{\sfL}{\mathsf{L}}

\begin{document}
\title{A nearly-$4\log n$ depth lower bound for formulas with restriction on top}%
\author{Hao Wu\thanks{College of Information Engineering, Shanghai Maritime University, Shanghai, China. My email is \texttt{haowu@shmtu.edu.cn}, you can also reach me via \texttt{ wealk@outlook.com}.}}%
\maketitle
\begin{abstract}
One of the major open problems in complexity theory is to demonstrate an explicit function which requires super logarithmic depth, a.k.a, the $\mathbf{P}$ versus $\mathbf{NC^1}$ problem.  The current best depth lower bound is $(3-o(1))\cdot \log n$, and it is widely open how to prove a super-$3\log n$ depth lower bound. Recently Mihajlin and Sofronova (CCC'22) show if considering formulas with restriction on top, we can break the $3\log n$ barrier. Formally, they prove there exist two functions $f:\{0,1\}^n \rightarrow \{0,1\},g:\{0,1\}^n \rightarrow \{0,1\}^n$, such that for any constant $0<\alpha<0.4$ and constant $0<\epsilon<\alpha/2$,  their XOR composition $f(g(x)\oplus y)$ is not computable by an AND of $2^{(\alpha-\epsilon)n}$ formulas of size at most $2^{(1-\alpha/2-\epsilon)n}$. This implies a  modified version of  Andreev function  is not computable by any circuit of depth $(3.2-\epsilon)\log n$ with the restriction that top $0.4-\epsilon$ layers only consist of AND gates for any small constant $\epsilon>0$.
They ask whether the parameter $\alpha$ can be push up to nearly $1$ thus implying a nearly-$3.5\log n$ depth lower bound.
 
In this paper, we provide a stronger answer to their question.  We show there exist two functions $f:\{0,1\}^n \rightarrow \{0,1\},g:\{0,1\}^n \rightarrow \{0,1\}^n$, such that for any constant $0<\alpha<2-o(1)$, their XOR composition $f(g(x)\oplus y)$ is not computable by an AND of $2^{\alpha n}$ formulas of size at most $2^{(1-\alpha/2-o(1))n}$. This implies a $(4-o(1))\log n$ depth lower bound with the restriction that top $2-o(1)$ layers only consist of AND gates. We prove it by observing that one crucial component in Mihajlin and Sofronova's work, called the well-mixed set of functions, can be significantly simplified thus improved. Then with this observation and a more careful analysis, we obtain these nearly tight results.

\end{abstract}

\newpage
\tableofcontents
\newpage

\section{Introduction}
One of the major open problems in complexity theory is to demonstrate an explicit function which requires super logarithmic depth, a.k.a, the $\mathbf{P}$ versus $\mathbf{NC^1}$ problem.  The current best depth lower bound \cite{DBLP:journals/siamcomp/Hastad98,DBLP:conf/focs/Tal14,DBLP:journals/eccc/Tal16a} is $(3-o(1))\cdot \log n$, and we still don't even know how to obtain  a lower bound strictly larger than $3\log n$. One promising approach to tackle this problem was suggested by  Karchmer, Raz and Wigderson \cite{DBLP:journals/cc/KarchmerRW95}, they proposed that we should understand the complexity of  (block)-composition of Boolean functions. Given two  functions $f: \{0,1\}^m \rightarrow  \{0,1\}$, $g: \{0,1\}^n \rightarrow  \{0,1\}$, we define their composite function $f\diamond g:(\{0,1\}^n)^m \rightarrow  \{0,1\}$ as:
$
f \diamond g\left(x_{1}, \ldots, x_{m}\right)=f\left(g\left(x_{1}\right), \ldots, g\left(x_{m}\right)\right).
$
Given any Boolean function $f$, we denote the depth complexity of $f$ by $\mathsf{D}(f)$, that is  the minimal depth of a circuit of AND, OR and NOT gates of fan-in $2$ that computes $f$.  And it is easy to see the depth complexity of  $f \diamond g$  is upper-bounded by $\mathsf{D}(f)+\mathsf{D}(g)$ and it is natural to ask whether the depth complexity of  $f \diamond g$  is far from this upper bound. Karchmer, Raz and Wigderson \cite{DBLP:journals/cc/KarchmerRW95} conjectured that the depth complexity of  $f \diamond g$  is not far from its upper bound:
\begin{conj}Given two arbitrary non-constant Boolean functions $f:\{0,1\}^{m} \rightarrow\{0,1\}$ and $g:\{0,1\}^{n} \rightarrow\{0,1\}$, then
	$
	\mathsf{D}(f \diamond g) \approx \mathsf{D}(f)+\mathsf{D}(g).
	$
\end{conj}
If the conjecture is proved and the ``approximate equality'' is instantiated with proper parameters,  by an argument of  iterative composition \cite{DBLP:journals/cc/KarchmerRW95}, we will obtain an explicit function with super-logarithmic depth, which separates $\mathbf{P} $ from $\mathbf{NC^1}$.  Many restricted cases of KRW conjecture have been proved to be true. For example, there are composition theorems when the inner function $g$ satisfies certain property \cite{DBLP:journals/siamcomp/Hastad98,DBLP:conf/focs/Tal14,DBLP:journals/cc/DinurM18,DBLP:conf/innovations/FilmusMT21}. There are composition theorems about universal relation \cite{DBLP:journals/cc/EdmondsIRS01,HW93,DBLP:journals/siamcomp/GavinskyMWW17,Koroth2018,DBLP:conf/coco/MihajlinS21,DBLP:journals/eccc/Wu23}. There are composition theorems where the composition itself is restricted such as monotone composition, semi-monotone composition \cite{DBLP:conf/focs/RezendeMNPR20} and strong composition \cite{DBLP:journals/eccc/Meir23}. There are also some variants \cite{DBLP:journals/cc/EdmondsIRS01,DBLP:journals/cc/Meir20,DBLP:conf/coco/MihajlinS21} of original conjecture with the similar effect to the $\mathbf{P}$ versus $\mathbf{NC^1}$ problem, but we don't know how to prove them either. Maybe to prove the general form of KRW conjecture is far out of our reach now.
 
Note that we don't even know how to prove a super-$3\log n$ depth lower bound, maybe we should consider following weaker conjecture which suffices to break the $3\log n$ barrier in the first place.
\begin{conj}\label{conj:1.2}There exist two non-constant Boolean functions $f,g:\{0,1\}^{n} \rightarrow\{0,1\}$ such that
$\mathsf{D}(f \diamond g) \geq {(1+\epsilon)n}$ for some small constant $\epsilon \in (0,1)$.
\end{conj}
Unfortunately, we don't even know how to prove this weaker conjecture. Currently, the closest answer to Conjecture \ref{conj:1.2}  is  Meir's strong composition theorem \cite{DBLP:journals/eccc/Meir23}, but we don't know how to prove it for the case of standard composition. Mihajlin and Sofronova \cite{DBLP:conf/coco/MihajlinS22} proposed we should consider proving depth lower bound against even weaker formulas by considering restriction on top of the formulas.  They managed to prove  a  composition  theorem for formulas with restriction on top via XOR composition. The so-called XOR composition, proposed by Mihajlin and Smal \cite{DBLP:conf/coco/MihajlinS21}, is a useful special case of standard composition. In fact, the first nontrivial composition theorem \cite{DBLP:conf/coco/MihajlinS21} of a universal relation and some function is proved via XOR composition.

Given two functions $f:\{0,1\}^n \rightarrow \{0,1\},g:\{0,1\}^n \rightarrow \{0,1\}^n$, their XOR composition $f\boxplus g$ is defined as :
 \[f\boxplus g(x,y)=f(g(x)\oplus y)\]
 where $\oplus$ denotes the bit-wise XOR of two binary strings. Mihajlin and Sofronova \cite{DBLP:conf/coco/MihajlinS22} proved following result.
\begin{thm}[\cite{DBLP:conf/coco/MihajlinS22}] \label{thm:1.3} If we choose a function $f:\{0,1\}^n \rightarrow \{0,1\}$ randomly, with probability $1-o(1)$, there exists a function $g:\{0,1\}^n \rightarrow \{0,1\}^n$, such that for any constant $0<\alpha<0.4$ and constant $0<\epsilon<\alpha/2$,  their XOR composition $f\boxplus g$ is not computable by an AND (or OR) of $2^{(\alpha-\epsilon)n}$ formulas of size at most $2^{(1-\alpha/2-\epsilon)n}$.
\end{thm}

This implies a super-$3\log n$ depth lower bound for a modified version of the Andreev function against formulas with restriction on top.
\begin{thm}[\cite{DBLP:conf/coco/MihajlinS22}]\label{thm:1.4} A modified version of  Andreev function $\mathbf{Andr}'$ is not computable by any circuit of depth $(3.2-\epsilon)\log n$ with the restriction that top $0.4-\epsilon$ layers only consist of AND(or OR) gates for any small constant $\epsilon>0$.
\end{thm}
Such kind of super-$3\log n$ depth lower bound is unknown before their work even for such strong restriction. They asked whether their result can be improved as asked by following question.
\begin{ques}[\cite{DBLP:conf/coco/MihajlinS22}]\label{ques:1.5}Is it possible to extend the  range of parameter $\alpha$ in Theorem \ref{thm:1.3}  to $0<\alpha <1$?
\end{ques}
In this paper, we give a positive answer to their question  with an even better result. In fact, we extend the  range of parameter $\alpha$ to $0<\alpha <2-o(1)$ which is nearly optimal. 
\subsection{Our results}
Our main result is an improved XOR composition theorem for formulas with restriction on top. Formally, we have following result.
\begin{thm}\label{thm:1.6}Let $\sfL(f)$ be the protocol size of any Boolean function $f$. For most  functions $f:\{0,1\}^n \rightarrow \{0,1\}$,  
	there exists a function $g : \{0, 1\}^n \rightarrow \{0, 1\}^n$, such that $f\boxplus g$ is not computable by an AND (or OR) of $2^{\alpha n}$ formulas of size at most  $2^{\frac{n+\log\mathsf{L}(f)-\alpha n-2\log\log n}{2}}$ for any $0<\alpha < 1+ \frac{\log\mathsf{L}(f)-2\log\log n}{n}$.
\end{thm}
This implies a nearly-$4\log n$ depth lower bound for formulas with restriction on top.
\begin{thm}\label{thm:1.7}A modified version of  Andreev function $\mathbf{Andr}'$ is not computable by any circuit of depth $(4-o(1))\log n$ with the restriction that top $(2-o(1))\log n$ layers only consist of AND (or OR) gates.
\end{thm}Comparing to the results of Mihajlin and Sofronova \cite{DBLP:conf/coco/MihajlinS22}, our results are nearly tight and our approach is much simpler to be described next.
\paragraph{Our approach.}Here we give a concise description of the proof idea of Theorem \ref{thm:1.6}. The whole proof strategy is similar to that in \cite{DBLP:conf/coco/MihajlinS22}, we call such strategy as a double-measurement argument, a generalized form of the double-counting argument. One crucial component in such argument is the notion of  well-mixed set of functions, and our improvement is mainly due to the simplification and improvement for such well-mixed set of functions. 

Let $\mathcal{G}$ be the set of all functions from $\{0,1\}^n \rightarrow \{0,1\}^n$. Now given a hard function $f:\{0,1\}^n \rightarrow \{0,1\}$, we want to show 
there exists a function $g \in \mathcal{G}$, such that if $f\boxplus g$ can be computed by a formula $\phi_g =\bigwedge_{i=1}^{2^{\alpha n}}
\phi_{g,i}$, there must be a sub-formula $\phi_{g,i}$ such that $\sfL(\phi_{g,i})$ is large. To show this, Mihajlin and Sofronova defined a sub-additive measure $\mu$ for Boolean functions of two arguments and $\mu(f\boxplus g)$ is large, thus by averaging, for every $g$, there exists some $i_g$ such that $\mu(\phi_{g,i_g})$ is large enough. Note that for every $g,i_g$, $\phi_{g,i_g}$ computes some function  $h_{g,i_g}$, and let  $\mathcal{H}$ be the set of all such functions $h_{g,i_g}$. If the size of $\mathcal{H}$ is large,  by a standard counting argument, there must be a  hard function $h \in \mathcal{H}$ as required. But $\mathcal{H}$ may be a small set, to prevent this, we need to show for every $h\in \mathcal{H}$, there  only exists a small subset of $\mathcal{G}_h \subseteq \mathcal{G}$ such that for every $g\in  \mathcal{G}_h$,  $h_{g,i_g}$ is the same function as $h$.  Formally, denote the $\{g|g\in \mathcal{G}, h_{g,i_g} =h\}$ by $\mathcal{G}_h$.
Let $h_\star$ be the function such that the size of $\mathcal{G}_{h_\star}$ is maximum among all  $\mathcal{G}_h$, it suffices to show $\mathcal{G}_{h_\star}$ is a small set. To this end, we need to show an up bound of measurement $\mu(h_\star)$ in another way and this is why the notion of well-mixed set is involved. 

Now consider this function 
$\mathcal{M}_{f,\mathcal{G}_{h_\star}}(x,y) =\bigvee_{g\in \mathcal{G}_{h_\star}} f(g(x)\oplus y)$. Let $\mathcal{M}_{f,\mathcal{G}_{h_\star}}^x$ be the function $\mathcal{M}_{f,\mathcal{G}_{h_\star}}$ with the first argument is fixed to be some $x\in \{0,1\}^n$, we want to show if the set $\mathcal{G}_{h_\star}$ is large, there are many $x$s such that $\mathcal{M}_{f,\mathcal{G}_{h_\star}}^x$ is almost a constant function, and it eventually  implies  $\mu(h_\star)$ is small which contradicts  the fact that $\mu(h_\star)$ is already large. This is essentially the property that Mihajlin and Sofronova wanted for $\mathcal{G}_{h_\star}$, or in their terms,  $\mathcal{G}_{h_\star}$ is well-mixed for function $f$. In their work, Mihajlin and Sofronova used a rather complicated probabilistic method to show that property.

We will show such complication is entirely unnecessary and the well-mixed property could be obtained by a simple counting argument if you choose the function $f$ properly. For convenience, let $N=2^n$,  now choose  a hard function $f:\{0,1\}^n \rightarrow \{0,1\}$ such that $\text{density}(f^{-1}(1)) = \frac{|f^{-1}(1)|}{N}\geq \delta$, typically, we set $\delta$ to be $\frac{1}{4}$. Note that given any fixed $x$, $\mathcal{M}_{f,\mathcal{G}_{h_\star}}^x(y) =1$ if there is a function $g\in \mathcal{G}_{h_\star}$ such that $(g(x)\oplus y) \in f^{-1}(1)$. Given any $x\in \{0,1\}^n$, denote the set $\{z|\exists g \in \mathcal{G}_{h_\star}, g(x)=z\}$ by $\mathcal{G}_{h_\star}(x)$. Similarly, given $x,y\in \{0,1\}^n$, we denote the set $\{z|\exists g \in \mathcal{G}_{h_\star}, g(x)\oplus y=z\}$ by $\mathcal{G}_{h_\star}(x)\oplus y$. Given an $x$, if $|\mathcal{G}_{h_\star}(x)| > N(1-\delta)$, for any fixed $y$, we also have $|\mathcal{G}_{h_\star}(x)\oplus y| > N (1-\delta)$, since for any fixed $y$, $z\oplus y$ is permutation function of $z$. When $|\mathcal{G}_{h_\star}(x)\oplus y| > N (1-\delta)$, $(\mathcal{G}_{h_\star}(x)\oplus y) \cap f^{-1}(1)$ is not empty,  this means there exists $g \in \mathcal{G}_{h_\star}$ such that for that $x$, $f(g(x)\oplus y)=1$.

 Now we say $x$ is bad, if $|\mathcal{G}_{h_\star}(x)| \leq N (1-\delta)$. If $\mathcal{G}_{h_\star}$ is a dense subset of $\mathcal{G}$, the number of bad $x$s is small. Assume the size of $\mathcal{G}_{h_\star}$ is (at least) $|\mathcal{G}| \cdot(1-\delta)^P$, then  there are at most $P$ bad $x$s. If not,  the number of functions in $\mathcal{G}_{h_\star}$ is less than \[\left(N(1-\delta)\right)^P \cdot N^{N-P} ={N^N} \cdot (1-\delta)^P =|\mathcal{G}| \cdot(1-\delta)^P\] which is a contradiction. This means, given any $x$ which is not bad, the function  $\mathcal{M}_{f,\mathcal{G}_{h_\star}}^x(y) =1$ for any $y$, thus $\mathcal{M}_{f,\mathcal{G}_{h_\star}}^x$ is a constant function. This eventually  implies  $\mu(h_\star)$ is small which leads to the desired contradiction. Finally, since $\mathcal{G}_{h_\star}$ has to be a small set, the set $\mathcal{H}$ must be a large set of functions which contains a hard function $h\in\mathcal{H}$ as required. See more details in Lemma \ref{lem:3.2} and Theorem \ref{thm:3.3}.

\paragraph{Other related works.}Besides Mihajlin and Sofronova's work, we note that Bathie and Williams \cite{bathie2024towards}   established a super-$3 \log n$ depth lower bound against \emph{uniform} circuits  consisting of only NAND gates. Since their result is against uniform circuits, it is incomparable to ours. They also pointed out that results similar to Mihajlin and Sofronova's work could be obtained from average-case lower bounds \cite{komargodski2017improved} via restriction-based techniques but they didn't provide further details. In a personal communication, Meir \cite{Meir24} pointed out that results similar to Mihajlin and Sofronova's work can also be obtained via techniques from communication complexity but it is not clear whether such results are as tight as ours.
\
\subsection{Organization of the rest of the paper}
The rest of the paper is organized as follows. In Section~\ref{sec:2}, we provide necessary preliminaries. In Section~\ref{sec:3}, we prove Theorem~\ref{thm:1.6}, an improved XOR composition theorem of formulas with restriction on top. In Section~\ref{sec:4}, we prove Theorem~\ref{thm:1.7}, a nearly-$4\log n$ depth lower bound for formulas with restriction on top. In Section~\ref{sec:5}, we conclude and make some discussion  about future directions.

\section{Preliminaries}\label{sec:2}
\begin{defn}[De Morgan formula]A  De Morgan  formula $\phi$ is a binary tree, its internal vertices are gates such as $\mathrm{AND}(\wedge)$ or $\mathrm{OR}(\vee)$, its leaves are  literals such as  $x_{i}$ or its negation $\neg x_{i}$. The depth of a formula is the depth of underling tree of the formula. The size of a formula is the number of its leaves. 
\end{defn}

\begin{defn}The formula complexity of a boolean function $f:\{0,1\}^{n} \rightarrow\{0,1\},$ denoted $\sfL(f),$ is the size of the smallest formula that computes $f .$ The depth complexity of $f,$ denoted $\mathsf{D}(f),$ is the smallest depth of a formula that computes $f$.
\end{defn}
We will need following fact, given a large set of distinct functions, there is a function with large formula size in that set.
\begin{fact}[\cite{DBLP:books/daglib/0028687}, Theorem 1.23]\label{fact:2.3}Let $\mathcal{F}$ be a set of distinct Boolean functions with input length $n$, then there exists a function $f\in \mathcal{F}$ such that $\sfL(f) \geq \frac{\log|\mathcal{F}|}{{\log n}+4}$.
\end{fact}

\begin{defn}[XOR composition of two functions, \cite{DBLP:conf/coco/MihajlinS21}]Let $f:\{0,1\}^n \rightarrow \{0,1\}, g:\{0,1\}^n \rightarrow \{0,1\}^n$ be two functions, their XOR composition $f\boxplus g :\{0,1\}^n \times \{0,1\}^n \rightarrow \{0,1\}$ is defined as follows:
	\[f\boxplus g(x,y)=f(g(x)\oplus y)\]
where $\oplus$ denotes the bit-wise XOR of two binary strings.
\end{defn}

We recall a measure $\mu(h)$ of any Boolean function $h$ of two arguments defined in  \cite{DBLP:conf/coco/MihajlinS22}.
\begin{defn}[\cite{DBLP:conf/coco/MihajlinS22}]Let $h$ be a Boolean function of two arguments, given any fixed $x$ as the first argument, we define the function $h^x$ by setting $h^x (y)=h(x,y)$ and define
	\[	\mu(h) = \sum_{x\in X}	\mathsf{L}(h^x).
	\]
\end{defn}

\begin{fact}[\cite{DBLP:conf/coco/MihajlinS22}]$\mu (f\boxplus g) \geq 2^n \cdot \sfL(f)$ and for every $x$, $\sfL(h^x) \leq \sfL(h)$.
\end{fact}


The  measure $\mu$ is sub-additive in the following sense:
\begin{lem}[\cite{DBLP:conf/coco/MihajlinS22}]\label{lem:2.7}Let $h(x, y) = \circ(g_1, g_2,\ldots , g_k)(x, y)$, where $\circ$ is $\wedge$ or $\vee$. Then $\mu (h) \leq  \mu (g_1) +\ldots + \mu (g_k).$
\end{lem}

\paragraph{Matrix representation for a function of two arguments.}
For convenience, we follow a notation in \cite{DBLP:conf/coco/MihajlinS22} which treats a Boolean function of two arguments as a Boolean matrix.
\begin{defn}Set $X=\{0,1\}^n, Y=\{0,1\}^n$, given a function $h:X \times Y \rightarrow \{0,1\}$, define a corresponding matrix $\mathcal{M}_h$  such that
	\begin{itemize}
		\item the rows of $\mathcal{M}_h$  are indexed by $x \in X$ and the columns are indexed by $y
		\in Y$,
		\item and $\mathcal{M}_h(x,y) = h(x,y)$ for every $x,y$.
	\end{itemize}
Similarly, given two functions  $f:\{0,1\}^n \rightarrow \{0,1\}, g:\{0,1\}^n \rightarrow \{0,1\}^n$, define a matrix $\mathcal{M}_{f,g}$ such that 
  $\mathcal{M}_{f,g}(x,y) = f(g(x)\oplus y)$ for every $x,y$.
  
 Furthermore, given a function  $f:\{0,1\}^n \rightarrow \{0,1\}$ and a set $\mathcal{Z}$ of functions from $\{0,1\}^n \rightarrow \{0,1\}^n$, define a matrix $\mathcal{M}_{f,\mathcal{Z}}$ such that  for every $x,y$,
  $$\mathcal{M}_{f,\mathcal{Z}}(x,y) = \bigvee_{g\in \mathcal{Z}} f(g(x)\oplus y).$$
  Finally, given a subset of indexes $A\subseteq X$ for rows in  a matrix $\mathcal{M}$,  the matrix $\mathcal{M}^A$ is a sub-matrix of $\mathcal{M}$ restricted to rows indexed by $A$.
  \end{defn}

\paragraph{Concentration of measure}
\begin{thm}[Chernoff bound]Given  $n$ independent random variables  $X_1,\ldots, X_n$ which distribute in $\{0, 1\}$,  let $X =\sum_{i=1}^n X_i$ be their sum and	$\mathbb{E}[X] = \mu$. For any constant $\delta$ such that $0<\delta <1$, we have
	$$\Pr(X \geq (1+\delta)\mu)\leq e^{-\frac{\delta^2 \mu}{2+\delta}}$$
	and 
		$$\Pr(X \leq  (1-\delta)\mu)\leq e^{-\frac{\delta^2 \mu}{2}}.$$
\end{thm}
By Chernoff bound, we have following fact.
\begin{fact}\label{fact:2.10}Let $N=2^n$. If we choose a function $f:\{0,1\}^n \rightarrow \{0,1\}$ randomly, then $\Pr(\frac{|f^{-1}(1)|}{N}< \frac{1}{4}) \leq e^{-\Omega(N)}$ and $\Pr(\frac{|f^{-1}(0)|}{N}< \frac{1}{4}) \leq e^{-\Omega(N)}$.
\end{fact}

\section{An improved XOR composition theorem for  formulas with restriction on top}\label{sec:3}
In this section, we prove Theorem \ref{thm:1.6}. At first, let's recall the notion of well-mixed set of functions.
\begin{defn}[Well-mixed set of functions, \cite{DBLP:conf/coco/MihajlinS22}]
	A set of functions $\mathcal{G}$ from $\{0, 1\}^n \rightarrow\{0, 1\}^n$ is
	$(Q, D, P)$-well-mixed for $f$ if $\forall \mathcal{Z} \subseteq \mathcal{G}, |\mathcal{Z}| \geq Q$, there exist a set $K\subseteq  \{0, 1\}^n, |K| \leq  P$ , such
	that $\mathcal{M}^{X\setminus K}_{f,\mathcal{Z}}
	$ has no more than $D$ zeroes in total  where $\mathcal{M}_{f,\mathcal{Z}}(x,y) =\bigvee_{g\in \mathcal{Z}} f(g(x)\oplus y)$.	
\end{defn}
Now we show given any approximately balanced function $f$, the set $\mathcal{G}$ of all functions  $\{0,1\}^n \rightarrow \{0,1\}^n$ is already a well-mixed set of functions for $f$.

\begin{lem}\label{lem:3.2} Let $f:\{0,1\}^n \rightarrow \{0,1\}$ be a function,  $\mathcal{G}$ be the set of all functions $\{0,1\}^n \rightarrow \{0,1\}^n$. For convenience, let $N =2^n$ and $X =\{0,1\}^n$. Assume  $\text{density}(f^{-1}(1)) = \frac{|f^{-1}(1)|}{N}\geq \delta$,  then $\mathcal{G}$ is $({|\mathcal{G}|}\cdot (1-\delta)^P,0,P)$-well-mixed for $f$.   Particularly, let  $\mathcal{Z} \subseteq \mathcal{G}$ be a set of functions and $\text{density}(\mathcal{Z}) = \frac{|\mathcal{Z}|}{|\mathcal{G}|}$, there exists a set  $K\subseteq \{0,1\}^n$ such that $|K|= P \leq \frac{\log \text{density}(\mathcal{Z}) }{ \log (1-\delta)}$  and all entries in $\mathcal{M}^{X\setminus K}_{f,\mathcal{Z}}$ are  ones.
\end{lem}

\begin{proof} Let  $\mathcal{Z} \subseteq \mathcal{G}$ be a set of functions, given any $x\in \{0,1\}^n$, denote the set $\{z|\exists g \in \mathcal{Z}, g(x)=z\}$ by $\mathcal{Z}(x)$, and we say $x$ is bad if $|\mathcal{Z}(x)|\leq N(1-\delta)$. Similarly, given $x,y\in \{0,1\}^n$, we denote the set $\{z|\exists g \in \mathcal{Z}, g(x)\oplus y=z\}$ by $\mathcal{Z}(x)\oplus y$.
	
Now assume there are  $P$ bad $x$s, the number of functions in $\mathcal{Z}$ is at most \[\left(N(1-\delta)\right)^P \cdot N^{N-P} ={N^N} \cdot (1-\delta)^P =|\mathcal{G}| \cdot(1-\delta)^P.\]
This implies
$\text{density}(\mathcal{Z}) = \frac{|\mathcal{Z}|}{|\mathcal{G}|} \leq  (1-\delta)^P$, this means $P\leq \frac{\log \text{density}(\mathcal{Z}) }{ \log (1-\delta)}$.

Now we show when $x$ is not bad, for every $y\in \{0,1\}$,   $\mathcal{M}_{f,\mathcal{Z}}(x,y) =\bigvee_{g\in \mathcal{Z}} f(g(x)\oplus y) =1$. Since $|\mathcal{Z}(x)|>N(1-\delta)$, we have $|\mathcal{Z}(x)\oplus y|>N(1-\delta)$.  Since  $|\mathcal{Z}(x)\oplus y| > N (1-\delta)$, $(\mathcal{Z}(x)\oplus y)\cap f^{-1}(1)$ is not empty, there must be a $g \in \mathcal{Z}$ such that $g(x)\oplus y \in f^{-1}(1)$, thus $\bigvee_{g\in Z} f(g(x)\oplus y)$ must be $1$ as required.
\end{proof}
Now we are ready to prove Theorem \ref{thm:1.6} rephrased as follows. It is proved via a similar idea  in \cite{DBLP:conf/coco/MihajlinS22} and a more careful analysis.
\begin{thm}\label{thm:3.3}
	Let  $f:\{0, 1\}^n \rightarrow \{0, 1\}$ be a function and 
	$\text{density}(f^{-1}(1)) \geq\frac{1}{4}$, 
	there exists a function $g : \{0, 1\}^n \rightarrow \{0, 1\}^n$, such that $f\boxplus g$ is not computable by an AND of $2^{\alpha n}$ formulas of size at most  $2^{\frac{n+\log\mathsf{L}(f)-\alpha n-2\log\log n}{2}}$ for any $0<\alpha < 1+ \frac{\log\mathsf{L}(f)-2\log\log n}{n}$.
\end{thm}
\begin{proof}[Proof of Theorem \ref{thm:3.3}]We prove it by contradiction. Let $\mathcal{G}$ be the set of all functions $\{0,1\}^n \rightarrow \{0,1\}^n$. Assume the contrary that for all $g \in \mathcal{G}$ the XOR composition $f\boxplus g$ is computable by   AND of  small enough formulas. That is, for any $g\in \mathcal{G}$, there is a formula $\phi_g$ computing $f\boxplus g$ and $\phi_g$ is  of following form
 \[	\phi_g =\bigwedge_{i=1}^{2^{\alpha n}}
	\phi_{g,i} \]
	where the size of every $\phi_{g,i}$ is at most  $2^{\frac{n+\log\mathsf{L}(f)-\alpha n-2\log\log n}{2}}$. Now let $h_{g,i}$ be the function that $\phi_{g,i}$ computes, thus $\sfL(h_{g,i}) \leq 2^{\frac{n+\log\mathsf{L}(f)-\alpha n-2\log\log n}{2}}$ and $f\boxplus g$ can be represented as  $\bigwedge_{i=1}^{2^{\alpha n}}
	h_{g,i}$. 
	
Recall that for every $g\in \mathcal{G}$, $\mu (f\boxplus g) \geq 2^n \cdot \sfL(f).$ By Lemma \ref{lem:2.7},  there must be an $i_g \in [2^{\alpha n}]$ such that, the measure $\mu(h_{g,i_g})$ is large:
$$\mu(h_{g,i_g} ) \geq 2^{(1-\alpha)n}\cdot \mathsf{L}(f).$$

Now we collect all such functions $h_{g,i_g}$ and let  $\mathcal{H}$ be the set of all such functions $h_{g,i_g}$. We want to show that the size of $\mathcal{H}$ is large, thus by a standard counting argument, there must be a function $h \in \mathcal{H}$ which requires large formulas  which contradicts the hypothesis. 

Given any $h\in \mathcal{H}$,  denote the $\{g|g\in \mathcal{G}, h_{g,i_g} =h\}$ by $\mathcal{G}_h$.
Let $h_\star$ be the function such that the size of $\mathcal{G}_{h_\star}$ is maximum among all  $\mathcal{G}_h$. We will prove $$-4 \log \text{density}(\mathcal{G}_{h_\star}) \cdot \sfL(h_\star) \geq  \mu(h_{\star} ) \geq 2^{(1-\alpha)n} \cdot \sfL(f),$$
before proving this, let's show it indeed leads to the contradiction required. Recall by assumption $\sfL(h_{\star}) \leq 2^{\frac{n+\log\mathsf{L}(f)-\alpha n-2\log\log n}{2}}$, this means
$$\text{density}(\mathcal{G}_{h_\star}) \leq 2^{-2^{\frac{n+\log\mathsf{L}(f)-\alpha n+2\log\log n-4}{2}}}.$$
Now we are ready to lower bound the size of $|\mathcal{H}|$, that is the number of distinct functions in $\mathcal{H}$.
Since $|\mathcal{H}|\cdot |\mathcal{G}_{h_\star}| \geq |\mathcal{G}|$,
$$|\mathcal{H}| \geq \frac{1}{\text{density}(\mathcal{G}_{h_\star})} \geq 2^{2^{\frac{n+\log\mathsf{L}(f)-\alpha n+2\log\log n-4}{2}}}.$$
By Fact \ref{fact:2.3}, there exists an $h\in \mathcal{H}$ such that 
\begin{align*}
\sfL(h) &\geq \frac{2^{\frac{n+\log\mathsf{L}(f)-\alpha n+2\log\log n-4}{2}}}{{\log 2n} +4 } \\
&> 2^{\frac{n+\log\mathsf{L}(f)-\alpha n-2\log\log n}{2}},\text{when $n$ is large enough,}
\end{align*}
which is a contradiction to the assumption. Now we show $-4\log \text{density}(\mathcal{G}_{h_\star}) \cdot \sfL(h_\star) \geq  \mu(h_{\star} ) $ by considering the matrix $\mathcal{M}_{f,\mathcal{G}_{h_\star}}$. By Lemma \ref{lem:3.2}, 
\begin{itemize}
	\item there exists a set  $K\subseteq \{0,1\}^n$ such that
	\begin{align*}
	|K|  &\leq    \frac{\log \text{density}(\mathcal{G}_{h_\star}) }{\log (1-1/4)} \\
	&\leq -4\cdot \log \text{density}(\mathcal{G}_{h_\star}), \text{ since  $\log \frac{3}{4} \approx -0.415$ thus $ 1 <\frac{\log \frac{3}{4}}{-0.25}$ }. 
	\end{align*}
	\item  And all entries in $\mathcal{M}^{X\setminus K}_{f,\mathcal{G}_{h_\star}}$ are  ones.
\end{itemize}
Now we have following key fact about the function $h_\star$.
\begin{fact}\label{fact:3.4}For every $x,y \in \{0,1\}^n$, $\mathcal{M}_{f,\mathcal{G}_{h_\star}}(x,y)=1$ implies $h_\star(x,y)=1$.
\end{fact}
\begin{proof}Recall that $\mathcal{M}_{f,\mathcal{G}_{h_\star}}(x,y) =\bigvee_{g\in \mathcal{G}_{h_\star}} (f\boxplus g)(x,y)$. If $\mathcal{M}_{f,\mathcal{G}_{h_\star}}(x,y)=1$, there must be some $g\in \mathcal{G}_{h_\star}$ such that $(f\boxplus g)(x,y) = 1$. Furthermore, by assumption $f\boxplus g$ is simply  $\bigwedge_{i=1}^{2^{\alpha n}}
	h_{g,i}$ and $h_\star =h_{g,i}$ for some $i$, thus  $h_\star(x,y)$ must be $1$ as well. 
\end{proof}
By  Fact \ref{fact:3.4},  given any $x\in X\setminus K$, for every $y\in \{0,1\}^n$, $h_{\star}^x(y) = 1$ , in other words, $h_{\star}^x$ is a constant function and $\sfL(h_{\star}^x)=0$. Now we are ready to up bound  $\mu(h_\star)$ and have
	\begin{align*}
	\mu(h_\star)&=\sum_x L(h_\star^x)=\sum_{x\in K} L(h_\star^x)+\sum_{x\notin K} L(h_\star^x) \\
	&= \sum_{x\in K} L(h_\star^x)\\
	&\leq |K|\cdot \sfL(h_\star) \\
	& \leq -4 \cdot \log \text{density}(\mathcal{G}_{h_\star}) \cdot \sfL(h_\star).
	\end{align*}
	as required.
\end{proof}
\begin{rem}We want to point out the AND($\wedge$) gate can be replaced with OR($\vee$) gate and since we use the counting argument to show there exists a hard function in the set $\mathcal{H}$, this result can be extended to the case of formula over full binary basis.
\end{rem}

\section{A nearly-$4\log n$ depth lower bound for  formulas with restriction on top}\label{sec:4}
We will prove a depth lower bound for a modified Andreev function defined in \cite{DBLP:conf/coco/MihajlinS22}.
\begin{defn}[\cite{DBLP:conf/coco/MihajlinS22}]The modified Andreev function $\mathbf{Andr}':\{0,1\}^n\times \{0,1\}^{n\log n} \times \{0,1\}^{n\times 2\log n } \rightarrow \{0,1\}$ is defined as follows:
	\[\mathbf{Andr}'(\mathtt{TT}_f,\mathtt{TT}_g,x_1,\ldots, x_{2\log n}) =(f\boxplus g)\left(\oplus(x_1),\ldots, \oplus(x_{2\log n})\right)\]
where  $\mathtt{TT}_f$ is a truth table of some function $f$ from $\{0,1\}^{\log n} \rightarrow \{0,1\}$,$\mathtt{TT}_g$ is a truth table of some function $g$ from $\{0,1\}^{\log n} \rightarrow \{0,1\}^{\log n}$, for every $i\in [2\log n]$, $x_i$ is a binary string of length $n$ and $\oplus(\cdot)$ is the parity function.
\end{defn}
\begin{thm}\label{thm:4.2}There exist two parameters $\gamma =o(1),\epsilon=o(1)$,  for every constant $\alpha$ such that $0<\alpha < 2-\gamma$, the modified Andreev function $\mathbf{Andr}'$ is not computable by an AND of $n^{\alpha}$ formulas of size at most $n^{3-\alpha/2-\epsilon}$.
	
In terms of depth lower bound, the modified Andreev function $\mathbf{Andr}'$ is not computable by any circuit of depth $(3+\alpha/2-\epsilon)\log n$ with the restriction that top $\alpha$ layers only consist of AND gates where $0<\alpha < 2-\gamma$. Choose $\alpha =2-o(1)$ properly, Theorem \ref{thm:1.7} follows.
\end{thm}
\begin{rem}
	Note that the input length of the modified Andreev function $\mathbf{Andr}'$ is $n'=(3\log n +1)n$,  writing the results in terms of $n^\prime$ doesn't change them essentially. For example,  $(4-o(1))\log n > (4-o(1))\log \frac{n'}{4\log n} > (4-o(1))\log \frac{n'}{4\log n'}=(4-o(1)-\frac{\log\log  n'+2}{\log n'})\log {n'} =(4-o(1))\log {n'}$. Similarly, in the restriction for top gates, AND could be replaced by OR. 
\end{rem}
Theorem \ref{thm:4.2} is proved via the same idea from \cite{DBLP:conf/coco/MihajlinS22}, the only differences here are  details of  parameters. For completeness of this paper, we present the proof here. At first, we recall the standard notion of random restriction.
\begin{defn}[Restriction]Given a Boolean function $f : \{0,1\}^n \rightarrow \{0,1\}$, a restriction $\rho\in \{0,1,\ast\}^n$ to function $f$ is a vector of length $n$ and for every $i\in[n]$, $\rho_i$ is an element from $\{0,1,\ast\}$.  Define $f|_\rho$ to be  the function restricted according to $\rho$ as follows:
if $\rho_i$ is $\ast$  then the $i$-th input bit of $f$ is unfixed thus free to be $0$ or $1$; otherwise the $i$-th input bit of $f$ is fixed to be $\rho_i$.
\end{defn}
\begin{defn}[Random restriction]Given $0<p<1$, the random restriction $\text{R}_p$  randomly samples restrictions as follows: every $\rho_i$ is sampled independently such that $\Pr[\rho_i=\ast]=p$ and $\Pr[\rho_i=1]=\Pr[\rho_i=0]=\frac{1-p}{2}$.
\end{defn}
In the rest of this section, we will set $p=\frac{2\ln\log n}{n}$.  Mihajlin and Sofronova  proved following useful two lemmas about random restriction of the modified Andreev function implicitly in \cite{DBLP:conf/coco/MihajlinS22}.

\begin{lem}[Implicit in  \cite{DBLP:conf/coco/MihajlinS22}]\label{lem:4.5}Let $\mathbf{Andr}'_{f,g}$ be the $\mathbf{Andr}'$ function hardwired with two fixed functions $f,g$. Then with probability $1-o(1)$, the random restriction $\text{R}_p$ will turn $\mathbf{Andr}'_{f,g}$ into $f\boxplus g$.
\end{lem}

\begin{lem}[Implicit in  \cite{DBLP:conf/coco/MihajlinS22}]\label{lem:4.6}Let $\alpha,\beta$ be two parameters such that $0<\alpha <2, 2<\beta<3$. Let  $\phi$  be a formula of form $\bigwedge_{i=1}^{n^{\alpha}}
	\phi_{i}$ where the size of each $\phi_{i}$ is at most $n^{\beta}$. Then with probability $1-o(1)$, the random restriction $\text{R}_p$ will turn $\phi$ into a formula $\phi'$  of form $\bigwedge_{i=1}^{n^{\alpha}}\phi_{i}^{\prime}$ where the size of each $\phi_{i}^{\prime}$ is at most $n^{\beta -2+\delta}$  for some  $\delta=o(1)$.
\end{lem}
Now we show how to prove Theorem \ref{thm:4.2} with above two lemmas.
\begin{proof}[Proof of Theorem \ref{thm:4.2}]At first choose some function $f$ from $\{0,1\}^{\log n}\rightarrow \{0,1\}$  and make sure that $f$ is  the function with maximum protocol size and $\frac{|f^{-1}(1)|}{n}\geq \frac{1}{4}$. By Fact \ref{fact:2.3} and \ref{fact:2.10}, we have $\sfL(f)\geq \log\left((1-o(1))2^{2^{\log n}}\right) /(\log\log n+4) \geq n/2\log\log n$ when $n$ is large enough. By Theorem \ref{thm:3.3}, we have following fact.
\begin{fact}\label{fact:4.7}Let $f$ be the function chosen above, there exists a function $g : \{0, 1\}^{\log n} \rightarrow \{0, 1\}^{\log n}$, such that $f\boxplus g$ is not computable by an AND of $n^{\alpha}$ formulas of size at most  $n^{1-\alpha/2-\frac{2\log\log\log n}{\log n}}$ for any $0<\alpha < 2-\frac{4\log\log\log n}{\log n}$ when $n$ is large enough.
\end{fact}
Let $\delta$ be the same parameter from Lemma \ref{lem:4.6}, $\gamma=\frac{4\log\log\log n}{\log n}$  and $\epsilon=\delta+\gamma/2$. Choose some  $\alpha$ such that $0<\alpha <2-\gamma$ and set $\beta=3-\alpha/2-\epsilon$. Now assume  Theorem \ref{thm:4.2} is false, that is  $\mathbf{Andr}'$ is computable by formula $\phi$  of form $\bigwedge_{i=1}^{n^{\alpha}}
\phi_{i}$ where the size of each $\phi_{i}$ is at most $n^{\beta}$, and so is $\mathbf{Andr}'_{f,g}$. By Lemma \ref{lem:4.5} and  \ref{lem:4.6}, $f\boxplus g$ is computable by a formula $\phi'$  of form $\bigwedge_{i=1}^{n^{\beta}}\phi_{i}^{\prime}$ where the size of each $\phi_{i}^{\prime}$ is at most $$n^{\beta -2+\delta} =n^{1-\alpha/2-\epsilon+\delta}=n^{1-\alpha/2-\frac{2\log\log\log n}{\log n}}$$
which contradicts Fact \ref{fact:4.7}.
\end{proof}

\section{Conclusion and discussion}\label{sec:5}
In this paper, we obtain a nearly-tight XOR composition theorem for formulas with restriction on top and with this composition theorem we have a nearly-$4\log n$ depth lower bound for formulas with restriction on top. Intuitively, in such the depth lower bound,  we trade one unrestricted layer to nearly two layers of AND gates on top of the circuit and our trade-off is nearly optimal.

The next nature question as pointed out by Mihajlin and Sofronova \cite{DBLP:conf/coco/MihajlinS22} is to prove the case with $\mathbf{AC}^0$ formula on top. But it turns out even to prove the case of depth-2 formula  with unbounded fan-in is difficult.  The obstacle to extending current approach to the case of depth-2 formula on top is that we don't know how to  find the sub-formula $h_\star$ such that the measure $\mu(h_\star)$ is large enough meanwhile $h_\star$ is correlated with $f\boxplus g$  properly like that in Fact \ref{fact:3.4}. This difficulty  also appears in the approach via communication complexity \cite{Meir24}, since such result  shares the same feature with a composition theorem of a depth-2 formula and a De Morgan formula, but we don't even know how to prove a composition theorem of two depth-2 formulas in general. New ideas are needed, maybe we should try to prove a general composition theorem of two depth-2 formulas in the first place.
 \section*{Acknowledgments}
The author would like to thank Or Meir for sharing the idea about how to prove results similar to Mihajlin and Sofronova's work via techniques from communication complexity and other helpful discussions. The author would  also like to thank Pei Wu for suggesting the question of the composition of two depth-2 formulas and other helpful discussions.

\bibliographystyle{alpha}
\bibliography{../hw}

\newcommand{\etalchar}[1]{$^{#1}$}
\begin{thebibliography}{GMWW17}

\bibitem[BW24]{bathie2024towards}
Gabriel Bathie and R~Ryan Williams.
\newblock Towards stronger depth lower bounds.
\newblock In {\em 15th Innovations in Theoretical Computer Science Conference
  (ITCS 2024)}. Schloss-Dagstuhl-Leibniz Zentrum f{\"u}r Informatik, 2024.

\bibitem[DM18]{DBLP:journals/cc/DinurM18}
Irit Dinur and Or~Meir.
\newblock Toward the {KRW} composition conjecture: Cubic formula lower bounds
  via communication complexity.
\newblock {\em Comput. Complex.}, 27(3):375--462, 2018.

\bibitem[dRMN{\etalchar{+}}20]{DBLP:conf/focs/RezendeMNPR20}
Susanna~F. de~Rezende, Or~Meir, Jakob Nordstr{\"{o}}m, Toniann Pitassi, and
  Robert Robere.
\newblock {KRW} composition theorems via lifting.
\newblock In Sandy Irani, editor, {\em 61st {IEEE} Annual Symposium on
  Foundations of Computer Science, {FOCS} 2020, Durham, NC, USA, November
  16-19, 2020}, pages 43--49. {IEEE}, 2020.

\bibitem[EIRS01]{DBLP:journals/cc/EdmondsIRS01}
Jeff Edmonds, Russell Impagliazzo, Steven Rudich, and Jir{\'{\i}} Sgall.
\newblock Communication complexity towards lower bounds on circuit depth.
\newblock {\em Comput. Complex.}, 10(3):210--246, 2001.

\bibitem[FMT21]{DBLP:conf/innovations/FilmusMT21}
Yuval Filmus, Or~Meir, and Avishay Tal.
\newblock Shrinkage under random projections, and cubic formula lower bounds
  for {AC0} (extended abstract).
\newblock In James~R. Lee, editor, {\em 12th Innovations in Theoretical
  Computer Science Conference, {ITCS} 2021, January 6-8, 2021, Virtual
  Conference}, volume 185 of {\em LIPIcs}, pages 89:1--89:7. Schloss Dagstuhl -
  Leibniz-Zentrum f{\"{u}}r Informatik, 2021.

\bibitem[GMWW17]{DBLP:journals/siamcomp/GavinskyMWW17}
Dmitry Gavinsky, Or~Meir, Omri Weinstein, and Avi Wigderson.
\newblock Toward better formula lower bounds: The composition of a function and
  a universal relation.
\newblock {\em {SIAM} J. Comput.}, 46(1):114--131, 2017.

\bibitem[H{\aa}s98]{DBLP:journals/siamcomp/Hastad98}
Johan H{\aa}stad.
\newblock The shrinkage exponent of de morgan formulas is 2.
\newblock {\em {SIAM} J. Comput.}, 27(1):48--64, 1998.

\bibitem[HW93]{HW93}
Johan Håstad and Avi Wigderson.
\newblock Composition of the universal relation.
\newblock In {\em ADVANCES IN COMPUTATIONAL COMPLEXITY THEORY, AMS-DIMACS},
  1993.

\bibitem[Juk12]{DBLP:books/daglib/0028687}
Stasys Jukna.
\newblock {\em Boolean Function Complexity - Advances and Frontiers}, volume~27
  of {\em Algorithms and combinatorics}.
\newblock Springer, 2012.

\bibitem[KM18]{Koroth2018}
Sajin Koroth and Or~Meir.
\newblock {Improved composition theorems for functions and relations}.
\newblock {\em Leibniz International Proceedings in Informatics, LIPIcs},
  116(48):1--18, 2018.

\bibitem[KRT17]{komargodski2017improved}
Ilan Komargodski, Ran Raz, and Avishay Tal.
\newblock Improved average-case lower bounds for de morgan formula size:
  Matching worst-case lower bound.
\newblock {\em SIAM Journal on Computing}, 46(1):37--57, 2017.

\bibitem[KRW95]{DBLP:journals/cc/KarchmerRW95}
Mauricio Karchmer, Ran Raz, and Avi Wigderson.
\newblock Super-logarithmic depth lower bounds via the direct sum in
  communication complexity.
\newblock {\em Comput. Complex.}, 5(3/4):191--204, 1995.

\bibitem[Mei20]{DBLP:journals/cc/Meir20}
Or~Meir.
\newblock Toward better depth lower bounds: Two results on the multiplexor
  relation.
\newblock {\em Comput. Complex.}, 29(1):4, 2020.

\bibitem[Mei23]{DBLP:journals/eccc/Meir23}
Or~Meir.
\newblock Toward better depth lower bounds: {A} krw-like theorem for strong
  composition.
\newblock {\em Electron. Colloquium Comput. Complex.}, {TR23-078}, 2023.

\bibitem[Mei24]{Meir24}
Or~Meir.
\newblock Personal communication, 2024.

\bibitem[MS21]{DBLP:conf/coco/MihajlinS21}
Ivan Mihajlin and Alexander Smal.
\newblock Toward better depth lower bounds: The {XOR-KRW} conjecture.
\newblock In Valentine Kabanets, editor, {\em 36th Computational Complexity
  Conference, {CCC} 2021, July 20-23, 2021, Toronto, Ontario, Canada (Virtual
  Conference)}, volume 200 of {\em LIPIcs}, pages 38:1--38:24. Schloss Dagstuhl
  - Leibniz-Zentrum f{\"{u}}r Informatik, 2021.

\bibitem[MS22]{DBLP:conf/coco/MihajlinS22}
Ivan Mihajlin and Anastasia Sofronova.
\newblock A better-than-3log(n) depth lower bound for de morgan formulas with
  restrictions on top gates.
\newblock In Shachar Lovett, editor, {\em 37th Computational Complexity
  Conference, {CCC} 2022, July 20-23, 2022, Philadelphia, PA, {USA}}, volume
  234 of {\em LIPIcs}, pages 13:1--13:15. Schloss Dagstuhl - Leibniz-Zentrum
  f{\"{u}}r Informatik, 2022.

\bibitem[Tal14]{DBLP:conf/focs/Tal14}
Avishay Tal.
\newblock Shrinkage of de morgan formulae by spectral techniques.
\newblock In {\em 55th {IEEE} Annual Symposium on Foundations of Computer
  Science, {FOCS} 2014, Philadelphia, PA, USA, October 18-21, 2014}, pages
  551--560. {IEEE} Computer Society, 2014.

\bibitem[Tal16]{DBLP:journals/eccc/Tal16a}
Avishay Tal.
\newblock Computing requires larger formulas than approximating.
\newblock {\em Electron. Colloquium Comput. Complex.}, {TR16-179}, 2016.

\bibitem[Wu23]{DBLP:journals/eccc/Wu23}
Hao Wu.
\newblock An improved composition theorem of a universal relation and most
  functions via effective restriction.
\newblock {\em Electron. Colloquium Comput. Complex.}, {TR23-151}, 2023.

\end{thebibliography}

\end{document}